\documentclass[]{spie}
\usepackage[utf8]{inputenc}
\usepackage[T1]{fontenc}
\usepackage{amsmath, amssymb}
\usepackage{graphicx}
\usepackage{cite}
\usepackage{hyperref}
\usepackage{authblk}
\usepackage{graphicx}
\usepackage{subcaption}
\usepackage{ragged2e} 
\usepackage{tabularx} 
\usepackage{booktabs} 
\usepackage{caption} 
\usepackage{multirow} 
\usepackage{multicol} 
\usepackage{makecell}
\usepackage{float}
\usepackage{comment}
\usepackage{tikz}
\title{RISTRETTO: a comparative performance analysis of the unmodulated Pyramid wavefront sensor and the Zernike wavefront sensor}
\author[1]{Muskan Shinde}
\author[1]{Nicolas Blind}
\author[1]{Christophe Lovis}
\affil[1]{D\'epartement d'Astronomie, Universit\'e de Gen\`eve, Chemin Pegasi 51, CH-1290 Versoix, Switzerland}
\date{} 
\begin{document}
\maketitle

\begin{abstract}
\noindent
The RISTRETTO instrument, a proposed visible high-contrast, high-resolution spectrograph for the VLT, has the primary science goal of detecting reflected light from nearby exoplanets and characterizing their atmospheres. Specifically, it aims to atmospherically characterize Proxima b, our closest temperate rocky exoplanet, located $37 mas$ from its host star, corresponding to $2\lambda/D$ at $\lambda=750 nm$. To achieve this goal, a raw contrast of less than $10^{-4}$ at $2\lambda/D$ and a Strehl ratio greater than 70\% are required, necessitating an extreme adaptive optics system (XAO) for the spectrograph. To meet the performance requirements for RISTRETTO, high sensitivity to low-order wavefront aberrations and petal modes is essential. Therefore, unmodulated Pyramid wavefront sensors (PWFS) and Zernike wavefront sensors (ZWFS) are under consideration. However, these sensors exhibit non-linearities and have a limited dynamic range, requiring different strategies to optimize their performance. The dynamic range of the sensors increases at longer wavelengths. Thus, in this study, we compare the performance of the 3-sided unmodulated PWFS, the 4-sided unmodulated PWFS, and the Zerniike WFS at different wavelengths in the visible and near-infrared regime.

\end{abstract}


\section{Introduction}

The RISTRETTO  instrument is a high-contrast, high-resolution spectrograph proposed as a visitor instrument for the VLT \cite{chazelas2020ristretto, lovis2022ristretto}. It aims to detect the reflected light from nearby exoplanets in the visible wavelength regime and to characterize their atmospheres. Specifically, it is driven by the goal to characterize the atmosphere of Proxima b, our closest temperate rocky exoplanet, which is located $37  mas$ from its host star, corresponding to $2\lambda/D$ at $\lambda=750 nm$  for an 8m-class telescope. To achieve this goal, a raw contrast of less than $10^{-4}$ at $2\lambda/D$ is required. Therefore, the spectrograph will be equipped with an extreme adaptive optics system (XAO) followed by a Phase Induced Amplitude Apodizer \& Nuller (PIAAN) coronagraph. Performing a tolerances analysis on the PIAA led to requirements of the Strehl at $750 nm$ to be above 70\% , or the total wavefront error (WFE) below $70 nm$ RMS, and the low-order WFE, within 3 cycles, to be less than $10 nm$ RMS, including pupil fragmentation effects \cite{blind_2022a}. To meet these, high sensitivity to low-order wavefront aberrations and petal modes is needed. Specifically, the wavefront sensor (WFS) must exhibit both high sensitivity and low wavefront noise. This can be accomplished through the use of Fourier-filtering wavefront sensors (FFWFS). 

\begin{figure}[h]
  \centering
  \includegraphics[width=0.5\textwidth]{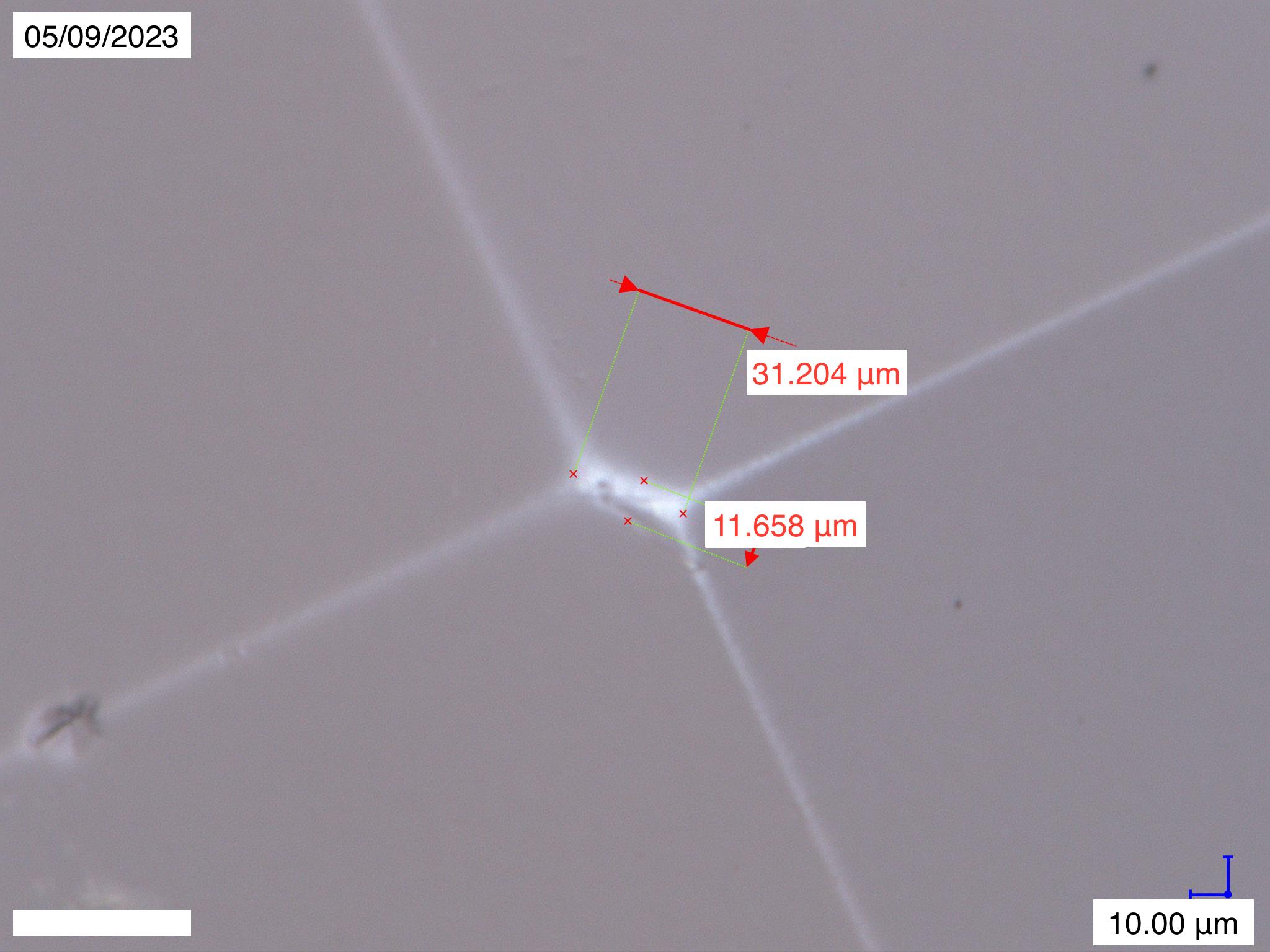}
  \caption{Roof apex created due to unequal polishing of the faces in 4PWFS (prototype kindly loaned by Laboratoire d'Astrophysique de Marseille).}
  \label{fig:roof_4PWFS}
\end{figure}

FFWFS is a class of WFS notable for their superior sensitivity. An FFWFS includes a phase mask positioned in an intermediate focal plane for optical Fourier filtering of the incoming wavefront \cite{fauvarque2016general}. A historical example of these Fourier-based WFSs is Foucault's knife-edge sensor from 1859 \cite{foucault1859memoire}. This concept was further generalized by Ragazzoni in 1996 with the PWFS \cite{ragazzoni1996pupil}. The PWFS comprises three main components: a tip-tilt mirror conjugated to the exit pupil of the system, a glass pyramid with its vertex at the focal point that splits light into the focal plane, and a relay lens that forms separated images of the exit pupil on a detector. When the tip-tilt mirror is not oscillating, the configuration is equivalent to a Foucault knife-edge test. However, with PWFS, it is possible to collect information on the x and y wavefront slopes simultaneously. The PWFS has gained considerable attention in recent years due to its high sensitivity and low wavefront noise, though it has a limited dynamic range and exhibits some non-linearities. To address these issues and increase the PWFS's linearity range, modulation is typically applied via an oscillating tip-tilt mirror tracing a circular pattern on the pyramid. This increases the effective spot size on the pyramid tip, enhancing the sensor’s linearity at the cost of sensitivity. The loop speed achievable with state-of-the-art tip-tilt mirrors barely reaches 1 kHz, but to meet RISTRETTO’s performance requirements, operation at least at 2 kHz is necessary \cite{blind_2024a}. This motivated us to move away from modulation, and towards the study of using an unmodulated PWFS, which could operate at high loop speeds and provide significant performance gains due to its high sensitivity. For the unmodulated PWFS, the quality of the pyramid tip is important. The classical pyramid used for this sensor is a 4-sided pyramid (4PWFS). Achieving the precesion of $0.2\lambda/D$ that we require at the tip of the 4PWFS is challenging even with significant effort in manufacturing pyramidal prisms by the companies we contacted. Unequal polishing depths of the pyramid faces often create a roof rather than a sharp tip as shown in Figure ~\ref{fig:roof_4PWFS}. As an alternative, a 3-sided PWFS (3PWFS) could be better, as it does not suffer from this roof defect. Due to lower number of pixels, it is also less sensitive to readout noise, which could be helpful for the high Strehl and high frame rates we need. Therefore, we are exploring a 3PWFS along with a 4PWFS.

Another type of FFWFS is the Zernike WFS (ZWFS), based on the Zernike phase-contrast concept initially proposed by Frits Zernike in the 1934 \cite{zernike1934diffraction}. It includes a phase-shifting dot, called the Zernike phase mask, placed at the focal plane, characterized by its diameter and its depth or phase shift. The mask introduces a phase change in the central part of the image. The light passing through the phase mask acts as a reference wavefront and interferes with the light passing outside the phase mask, which contains information about wavefront aberrations. The interfered light is reimaged onto the subsequent pupil plane, producing an intensity pattern related to the input wavefront aberration. The exact intensity encoding of the WFE depends on the phase difference and the size of the phase mask. The ZWFS is known for its high sensitivity, though it has an even more limited dynamic range compared to the PWFS, and is currently considered as a secondary WFS in existing AO systems.

In this study we will compare the performance of 3PWFS, 4PWFS and ZWFS. The mathematical description of the FFWFS and the transfer function computation to measure the sensitivity of the 3PWFS and 4PWFS is elaborated in Section II. Simulations to study the dynamic range of the 3PWFS, 4PWFS, and ZWFS at different wavelengths are presented in Section III. The study of their ability to detect petal modes is provided in Section IV. The optical gains are discussed in Section V. Conclusions are presented in Section VI.

\section{Sensitivity of FFWFS}

All the FFWFSs can be unified under the same mathematical formalism, allowing the development of a robust criterion for comparing their performance in the context of wavefront sensing. We use this formalism to establish the sensitivity function of the 3PWFS, and compare it to the 4PWFS. 

\begin{figure}[h]
  \centering
  \includegraphics[width=0.8\textwidth]{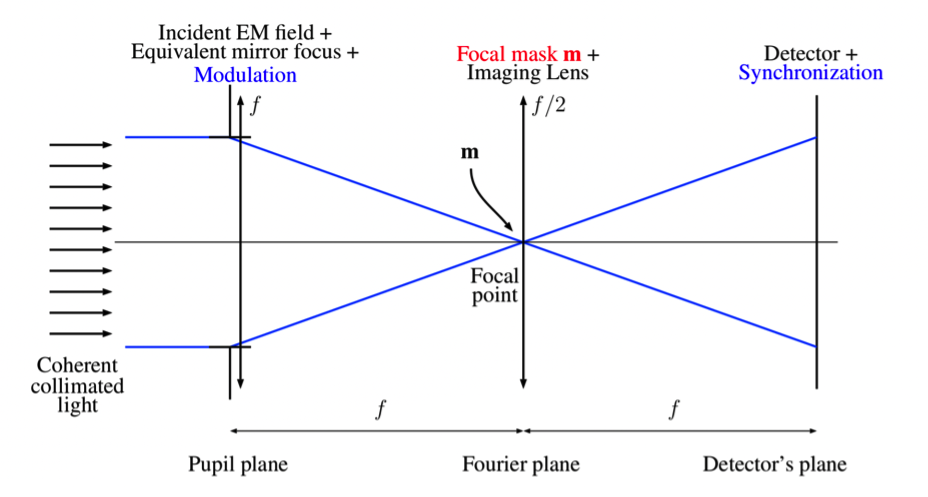}
  \caption{Schematic view of a Fourier filtering optical system. \cite{fauvarque2017general}}
  \label{fig:FFWFS_optical_design}
\end{figure}

\noindent
The general framework of the Fourier-based wavefront sensing to classify all FFWFSs given by Fauvarque et al. 2016 \cite{fauvarque2016general, fauvarque2017general} is described here. The optical design is shown in Figure ~\ref{fig:FFWFS_optical_design}. For a typical Fourier filtering system, the incoming electromagnetic field \(\Psi_p\) is described by:

\begin{equation} \label{eqn}
\Psi_p (\phi,\eta) = \sqrt{\eta} \, I_p \, e^{i\phi}
\end{equation}

\noindent
where \(\eta\) is the spatial-averaged flux, \(\phi\) is the phase at the analysis wavelength \(\lambda_0\), and \(I_p\) is the indicative function of the entrance pupil.

\noindent
The Fourier mask, which is placed at the focal point, is considered as a pure phase mask $m$:

\begin{equation} \label{eqn}
m = e^{\frac{2i\pi}{\lambda_0} \text{OS}}
\end{equation}

\noindent
where OS represents the optical shape of the mask. Using the Fraunhofer optical formalism, the intensity on the detector can be written as:

\begin{equation} \label{eqn}
I(\phi,\eta) = \left|\Psi_p (\phi,\eta) * F[m]\right|^2
\end{equation}

\noindent
where \(F[m]\) is the 2D Fourier transform of the mask. The phase seen by the WFS is the sum of the turbulent phase induced by the atmosphere and the static aberrations of the wavefront sensing path. It can be split into two terms:

\begin{equation} \label{eqn}
\phi = \phi_t + \phi_r
\end{equation}

\noindent
where \(\phi_t\) is the turbulent phase, and \(\phi_r\) is the static reference phase. \(\phi_r\) may also be seen as the operating point of the WFS. Using Taylor's expansion of the phase around the reference phase \(\phi_r\), the intensity on the detector can be written as a power series in phase:

\begin{equation} \label{eqn}
I(\phi,\eta) = \eta \left(I_c + I_l (\phi_t) + I_q (\phi_t) + \cdots \right)
\end{equation}

\noindent
The first term \(I_c\) is the constant intensity, which corresponds to the intensity on the detector when the phase equals the reference phase, i.e., when the turbulent phase is null. The second term \(I_l\) is the phase-linear term, which corresponds to the perfectly linear dependence of the intensity regarding the turbulent phase around the reference phase. The third term \(I_q\) is the quadratic intensity, which corresponds to the first non-linear dependence of the intensity. The next terms are non-linear contributions as well. The easiest way to create a phase-linear quantity from \(I(\phi,\eta)\) is to calculate the meta-intensity, called \(mI\), via the following equation:

\begin{equation} \label{eqn}
mI (\phi_t) = \frac{1}{\eta} \left(I(\phi_t + \phi_r,\eta) - I(\phi_r,\eta)\right)
\end{equation}

\noindent
The normalization by the factor \(\eta\) allows to make \(mI\) independent from the incoming flux. The constant reference intensity \(I(\phi_r,\eta)\) is computed via a calibration path. This ensures that the meta-intensity equals zero when there is no turbulent incoming phase. With such a definition, the meta-intensity \(mI\) equals the linear intensity \(I_l\) in the small phase approximation regime.

One of the crucial criteria to evaluate the performance of WFSs is their sensitivity. We would like to compute the sensitivity \(s(\phi_t)\) associated with a normalized turbulent mode \(\phi_t\). Which can be computed based on a convolutional model, as described in Fauvarque et al. 2019 \cite{fauvarque2019kernel}. This model assumes that the sensor can be fully characterized by an impulse response (IR) that links the entrance phase to the measured meta intensities.

\begin{equation} \label{eqn}
mI (\phi_t) = I_l (\phi_t) = I_p \phi_t * \text{IR}
\end{equation}

\noindent
where \(*\) is the convolutional product. A convenient aspect of the convolutional approach is the fact that we can compute the transfer function (TF) of an FFWFS. The TF is the 2D Fourier transform of the IR. It gives the response of the WFS in the spatial frequencies space, i.e., in a focal plane. In the focal plane, FFWFSs have a phase mask \(m\) and their weighting functions \(\omega\), which describe the energy distribution in the focal plane during one acquisition time of the sensor. This \(\omega\) can be seen as an effective modulation weighting function. \(\omega\) is normalized to 1 in order to ensure energy conservation. Assuming that \(\omega\) is a real function and is system around the centre. TF is expressed through the following formula \cite{fauvarque2019kernel}:

\begin{equation} \label{eqn}
\text{TF} = 2\text{Im}[m * (\bar{m}\omega)]
\end{equation}

\noindent
where \(\text{Im}\) is the imaginary part and the bar is the complex conjugate operator.

\noindent
The sensitivity with respect to spatial frequencies can be given by:

\begin{equation} \label{eqn}
s(\phi_t)_k = \sqrt{|TF|^2 * \text{PSF}}_k
\end{equation}

\noindent
where \(k\) can be seen as a position vector in the spatial frequencies space, PSF is the point spread function of the system given by \(|I_p|^2\).

\begin{figure}[H]
  \centering
  \begin{subfigure}[b]{0.2\textwidth}
    \centering
    \includegraphics[width=\textwidth]{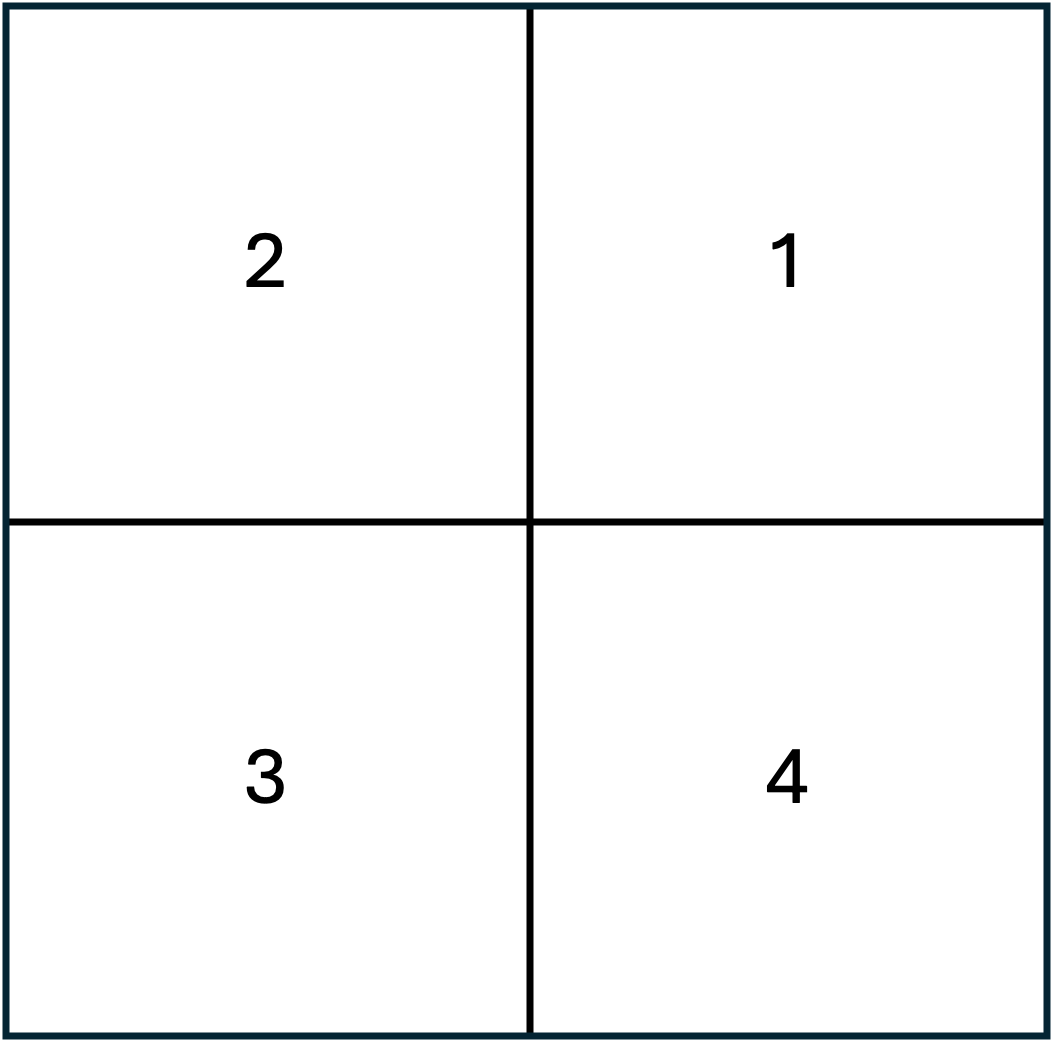}
    \caption{4PWFS}
  \end{subfigure}
  \hspace{0.03\textwidth} 
  \begin{subfigure}[b]{0.2\textwidth}
    \centering
    \includegraphics[width=\textwidth]{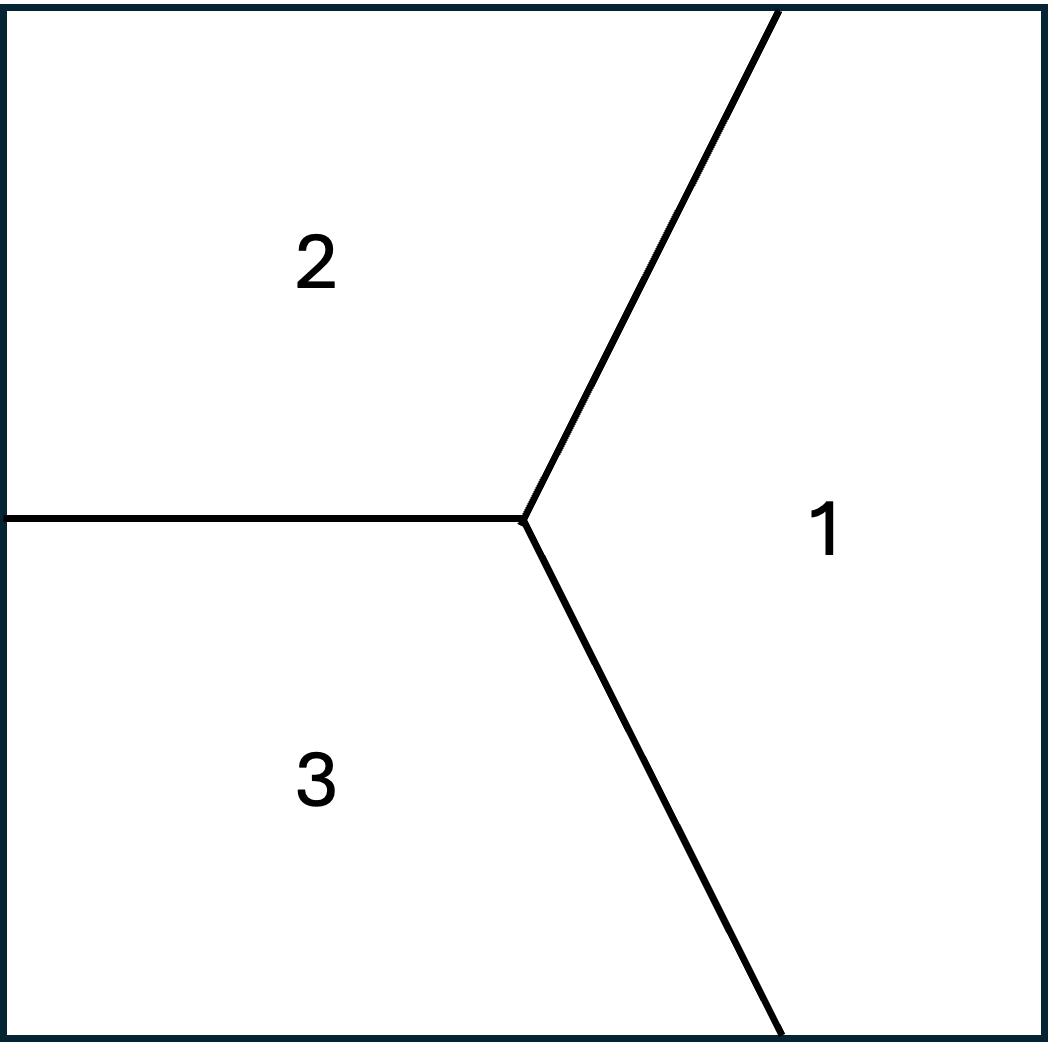}
    \caption{3PWFS}
  \end{subfigure}
  \caption{The tessellations of masks for 4PWFS and 3PWFS}
  \label{fig:testellations}
\end{figure}

We use the developments by Fauvarque et al. (2019) \cite{fauvarque2019kernel} for the PWFS. To handle independently the pupil images, we assume that they are generated by four Fourier-based WFSs using different filtering masks \(m_i\). Each of them corresponds to one of the pupil images created by the pyramid. The tessellation for phase masks for 4PWFS and 3PWFS is shown in Figure ~\ref{fig:testellations} . It is assumed that the modulation function \(\omega\) is identical for these four WFSs. The meta-intensity \(mI_i\) for each of the \(m_i\) is given by:

\begin{equation} \label{eqn}
mI_i (\phi_t) = I_p \phi_t * \text{IR} (m_i,\omega)
\end{equation}

\begin{equation} \label{eqn}
\text{IR} (m_i,\omega) = 2\text{Im}[\bar{F[m_i]} (F([m_i] * F[\omega])]
\end{equation}

\noindent
where F is the 2D Fourier transform operator.

\noindent
The classical way is to combine the signal to produce slopes map in the x and y directions. The slope maps for the 4PWFS are defined as:

\[
S_{x,4PWFS} = mI_1 (\phi_t) + mI_4 (\phi_t) - mI_2 (\phi_t) - mI_3 (\phi_t)
\]

\begin{equation} \label{eqn}
S_{y,4PWFS} = mI_1 (\phi_t) + mI_2 (\phi_t) - mI_3 (\phi_t) - mI_4 (\phi_t)
\end{equation}

\noindent
Whereas the slope maps for the 3PWFS are defined as \cite{schatz2021three}:

\[
S_{x,3PWFS} = mI_1 (\phi_t) - \frac{1}{2} mI_2 (\phi_t) - \frac{1}{2} mI_3 (\phi_t)
\]

\begin{equation} \label{eqn}
S_{y,3PWFS} = \sqrt{\frac{3}{2}} mI_2 (\phi_t) - \sqrt{\frac{3}{2}} mI_3 (\phi_t)
\end{equation}

\noindent
The slope maps help us understand physically how the pyramid performs the wavefront sensing. \(S_x\) and \(S_y\) can be seen as the phase derivative in the spatial frequency space along the x-axis and y-axis. Since slope maps are computed by linear transform of the meta intensities, it is possible to associate them with the IRs. For the 4PWFS, the IRs can be given by:

\[
IR_{x,4PWFS} = IR (m_1,\omega) + IR (m_4,\omega) - IR (m_2,\omega) - IR (m_3,\omega)
\]

\begin{equation} \label{eqn4pwfs}
IR_{y,4PWFS} = IR (m_1,\omega) + IR (m_2,\omega) - IR (m_3,\omega) - IR (m_4,\omega)
\end{equation}

\noindent
Whereas for the 3PWFS, the IRs can be given by:

\[
IR_{x,3PWFS} = IR (m_1,\omega) - \frac{1}{2} IR (m_2,\omega) - \frac{1}{2} IR (m_3,\omega)
\]

\begin{equation} \label{eqn3pwfs}
IR_{y,3PWFS} = \sqrt{\frac{3}{2}} IR (m_2,\omega) - \sqrt{\frac{3}{2}} IR (m_3,\omega)
\end{equation}

\noindent
The TFs can then be computed from the Fourier transform of IRs:

\[
TF_x = F[IR_x]
\]

\begin{equation} \label{eqn}
TF_y = F[IR_y]
\end{equation}

\begin{figure}[]
    \centering
    
    \begin{subfigure}[b]{0.495\textwidth}
        \centering
        \includegraphics[width=\textwidth]{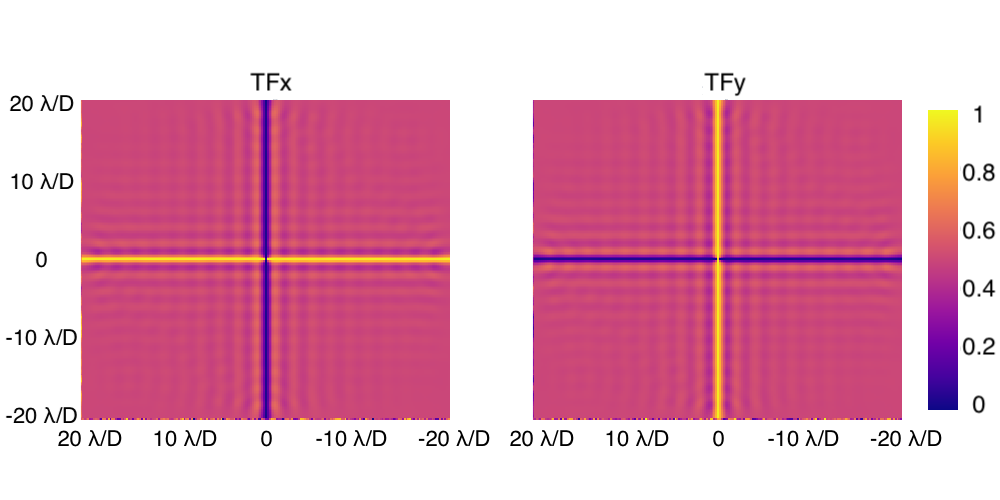}
        \caption{4PWFS convolutional}
    \end{subfigure}
    \hfill
    \begin{subfigure}[b]{0.495\textwidth}
        \centering
        \includegraphics[width=\textwidth]{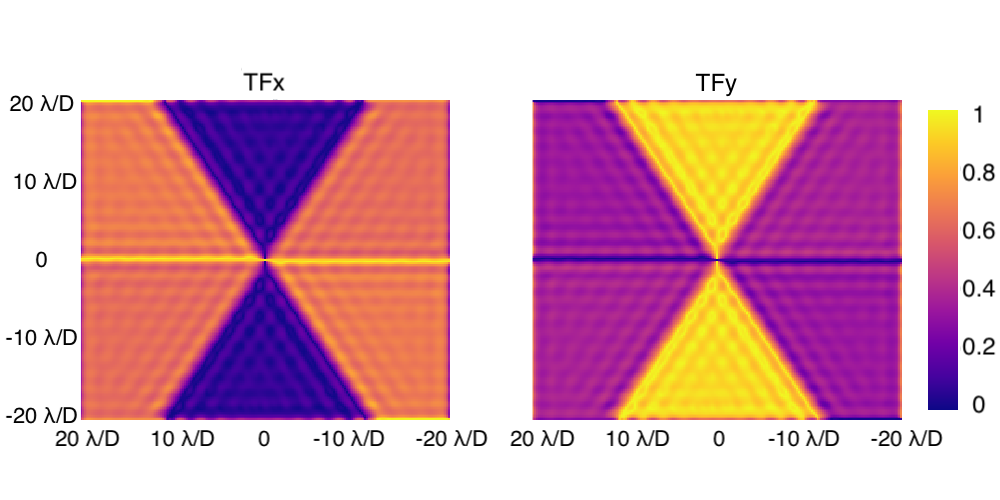}
        \caption{3PWFS convolutional}
    \end{subfigure}
    
    \begin{subfigure}[b]{0.495\textwidth}
        \centering
        \includegraphics[width=\textwidth]{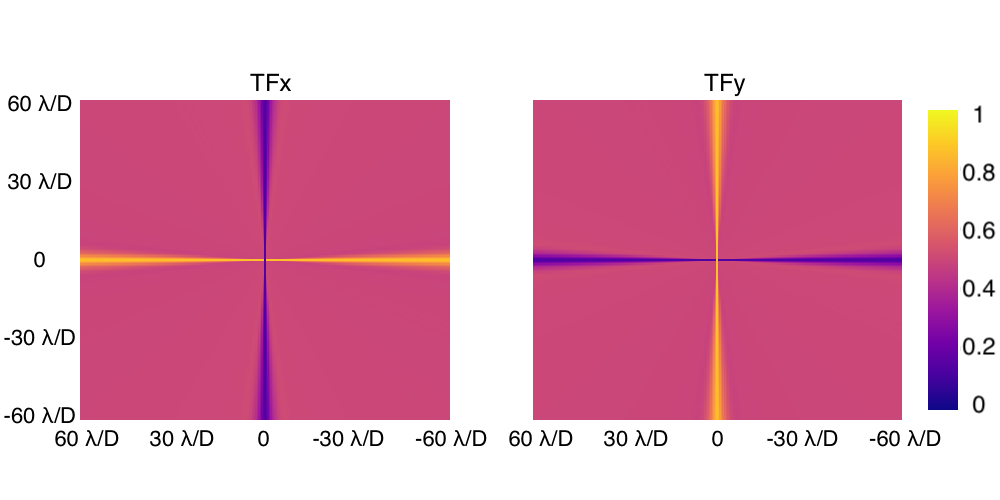}
        \caption{4PWFS end-to-end}
    \end{subfigure}
    \hfill
    \begin{subfigure}[b]{0.495\textwidth}
        \centering
        \includegraphics[width=\textwidth]{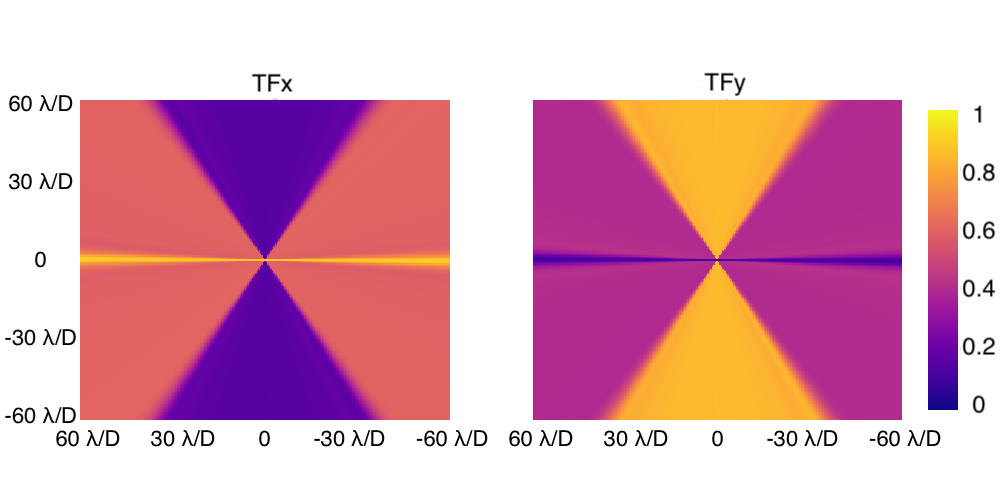}
        \caption{3PWFS end-to-end}
    \end{subfigure}

    \caption{Transfer Function TFx and TFy for the PWFS based (top) Convolution model and (bottom) End-to-end simulations for (left) 4PWFS and (right) 3PWFS.}
    \label{fig:TFs}
\end{figure}

\noindent
The TF\(_x\) and TF\(_y\) for the 4PWFS and 3PWFS are shown in Figure ~\ref{fig:TFs} (top).

Figure ~\ref{fig:TFs} (bottom) shows the TFs obtained from the full end-to-end physical optics model using the HCIpy Python library \cite{por2018hcipy}. We create a Fourier mode for a given spatial frequency in the focal plane. Then, we propagate this Fourier mode through the PWFS and record its response. Subsequently, we compute the slope maps \(S_x\) and \(S_y\) from the PWFS response. The RMS of the slope maps is stored as the TF value for that spatial frequency. This process is repeated for all the spatial frequencies in the focal plane, ranging from $-60\lambda/D$ to $60\lambda/D$ in both the u and v directions, to sample the entire focal plane.

In the case of the 4PWFS, for the TF\(_x\), we observe that the maximum sensitivity lies around the x-axis, while the least sensitivity is around the y-axis. This means that \(S_x\) measures the spatial frequencies in the x direction but is blind to those in the y direction. The TF\(_y\) is complementary to TF\(_x\), so the combination of the two slope maps can measure all the spatial frequencies. The spatial frequencies in the middle of these maps have degraded sensitivity in both slope maps, while the total sensitivity is maximum close to the edges of the pyramid mask. Similar results were observed in \cite{deo2018assessing} and \cite{correia2020performance}.

In the case of the 3PWFS, for the TF\(_x\), we observe that the maximum sensitivity lies around the x-axis, while the least sensitivity lies around a triangular region along the y-axis. For the TF\(_y\), the maximum sensitivity lies around a triangular region along the y-axis, with the least sensitivity along the x-axis. These triangular regions coincide with the mask considered for the 3PWFS, aligning with the orientation of the 3PWFS\textbf: the corresponding spatial frequencies do not reach face \#1, which only contains x-slope information (Equation~\ref{eqn3pwfs}). Since the definitions of x and y are related to the relative orientation of the pyramid to the detector, this "dead zones" are not a concern. Rotating one or the other rotates the x/y axis definition relative to the pyramid. In fact the same is true for the 4PWFS.


\section{Linearity}

Another key criterion for evaluating the performance of a WFS is its linearity or dynamic range. The WFS's linear behavior is encapsulated by an Interaction Matrix (IM), which is determined through a calibration process. 

\begin{figure}[H]
  \centering
  \includegraphics[width=0.48\textwidth]{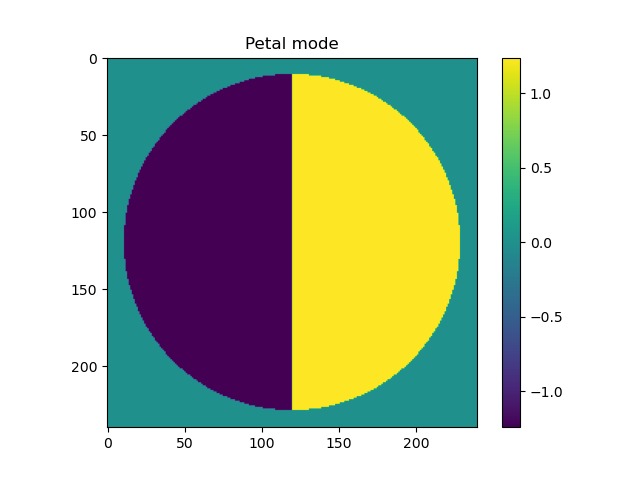}
  \caption{The step petal mode used in our simulations. The colorbar represents the phase amplitude in radians.}
  \label{fig:petal_mode}
\end{figure}

\begin{table}[H]
\centering
\caption{Simulation parameters}
\begin{tabular}{|l|l|}
\hline
\textbf{Telescope}         &                                \\ \hline
Pupil diam. {[}m{]}        & 8                              \\ \hline
Pupil samp. {[}pix{]}      & 240                            \\ \hline
Obs. ratio {[}\%{]}        & 0                              \\ \hline
\textbf{WFS}               &                                \\ \hline
No. sub-apertures          & $80 \times 80$                 \\ \hline
Sub-pupils separation {[}pix{]} & 40                           \\ \hline
Sensing wavelength {[}nm{]} & 600, 1000, 1600               \\ \hline
\textbf{DM}                &                                \\ \hline
No. actuators              & $40 \times 40$                 \\ \hline
No. modes                  & 1600                           \\ \hline
Modal basis                & Petal plus Zernike modes            \\ \hline
\textbf{Control}           &                                \\ \hline
Approach                   & Modal                          \\ \hline
Reconstruction matrix      & Singular value decomposition (SVD)          \\ \hline
\end{tabular}
\label{tab:simulation_parameters}
\end{table}

The intensities of the WFS's linear responses to a series of incoming phases \(\phi_i\), which together form the basis of the phase space we aim to control, are recorded. For each phase mode, the slopes of the linear response \(\delta I(\phi_i)\) are calculated using a push-pull approach:

\begin{equation} \label{eqn}
\delta I(\phi_i) = \frac{I(a\phi_i) - I(-a\phi_i)}{2a}
\end{equation}

\noindent
where \(I(a\phi_i)\) is the intensity recorded on the detector for a mode \(\phi_i\) and \(a\) is the amplitude of that mode. The IM is generated by concatenating slopes recorded for all modes in the basis, represented as:

\begin{equation} \label{eqn}
\text{IM} = (\delta I(\phi_1), \ldots, \delta I(\phi_i), \ldots, \delta I(\phi_N))
\end{equation}

\noindent
where N is the the total number of modes in the basis. The IM helps to associate the incoming phase with WFS measurements. To go from WFS measurements to wavefront phase, a Reconstruction Matrix (RM) is computed by performing a pseudo-inverse of the IM:

\begin{equation} \label{eqn}
\text{RM} = \text{IM}^{-1}
\end{equation}

\noindent
For a particular phase mode \(\phi_i\), the wavefront phase measured by the WFS can be calculated as:

\begin{equation} \label{eqn}
a(\phi_i) = (\text{RM} \cdot I(a\phi_i))_i
\end{equation}

We perform AO simulations to evaluate the dynamic range of the WFS. This is done using the end-to-end simulation tool we developed with the Hcipy Python library. The simulation parameters are provided in Table \ref{tab:simulation_parameters}.

\begin{figure}[H]
    \centering
    \begin{subfigure}{\textwidth}
        \centering
        \includegraphics[width=0.62\textwidth]{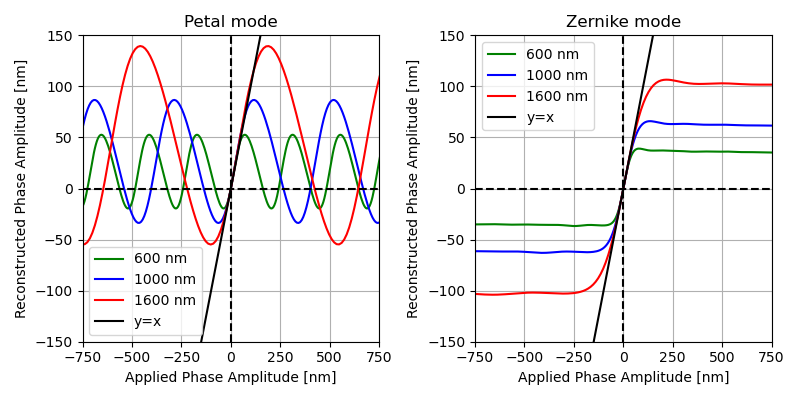}
        \caption{3PWFS}
        \label{subfig:3pwfs}
    \end{subfigure}
    
    \begin{subfigure}{\textwidth}
        \centering
        \includegraphics[width=0.62\textwidth]{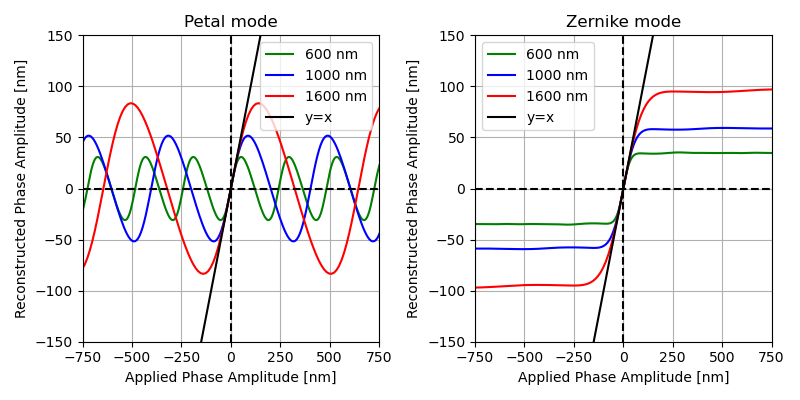}
        \caption{4PWFS}
        \label{subfig:4pwfs}
    \end{subfigure}
    
    \begin{subfigure}{\textwidth}
        \centering
        \includegraphics[width=0.62\textwidth]{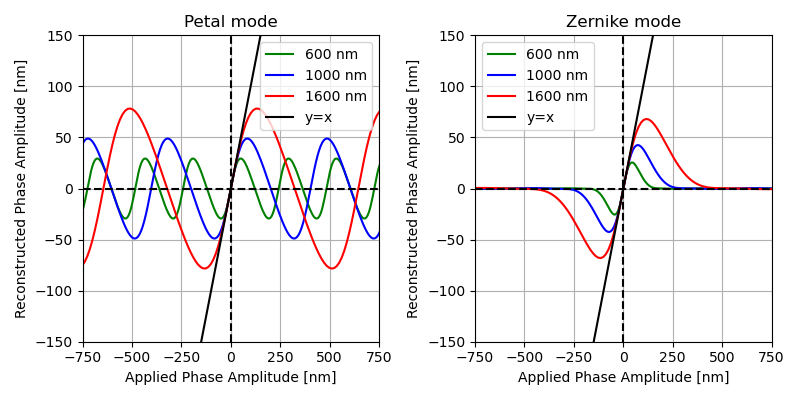}
        \caption{ZWFS}
        \label{subfig:zwfs}
    \end{subfigure}
    \caption{Linearity plots of (a) 3PWFS, (b) 4PWFS, and (c) ZWFS for petal mode (left) and Zernike mode (right) across wavelengths of 600 nm, 1000 nm, and 1600 nm at wavelenghts 0f 600nm, 1000nm, and 1600nm.}
    \label{fig:linearity_WFS}
\end{figure}

We use the Zernike basis for control, including the petal mode in the basis. The petal modes are piston jumps over each of the four segments of the pupil, generated due to the Low Wind Effect (LWE). The LWE occurs when the spiders holding the secondary mirror become significantly cooler than the surrounding air. This temperature difference, due to the radiative cooling of the telescope spiders, leads to an optical path difference between each side of the spider and causes phase discontinuities in the incoming wavefront at the location of the spiders \cite{sauvage2016tackling}. We used a simplified, 0-mean version for our simulations, as shown in Figure ~\ref{fig:petal_mode}.

The linearity curves at 3 distinct wavelengths for the 3PWFS, 4PWFS, and ZWFS for the petal modes and a Zernike mode is shown in Figure ~\ref{fig:linearity_WFS}.  It is evident that the dynamic range of all the WFSs increases by more than a factor of 2 when we go from 600 nm to 1600 nm, as expected. For Zernike modes, the PWFSs display a larger dynamic range in comparison to the ZWFS. At the wavelenght of 1600 nm, the PWFSs offer a dynamic range between 140-160 nm, while the ZWFS offers that of 100 nm. We also observe that the 3PWFS offers a slightly larger dynamic range than that of the 4PWFS, for the the 3PWFS it is around 20 nm larger than the 4PWFS. All three WFS are able to sense the petal modes, and their response to the petal mode is periodic. It is worth noting that, conversely to the 3PWFS, the 4PWFS has a symmetric response across zero for the petal mode we considered. This symmetry dispappears with more realistic petals or a different orientation of petal with respect to the 4PWFS.


\section{Petal modes sensing}
For RISTRETTO, we need to control petal modes to better than 10 nm PTP \cite{blind_2024a}. We perform AO simulations to understand how petal modes are measured and corrected by the AO system.

\begin{figure}[H]
    \centering
    
    \begin{subfigure}[b]{0.32\textwidth}
        \centering
        \includegraphics[width=\textwidth]{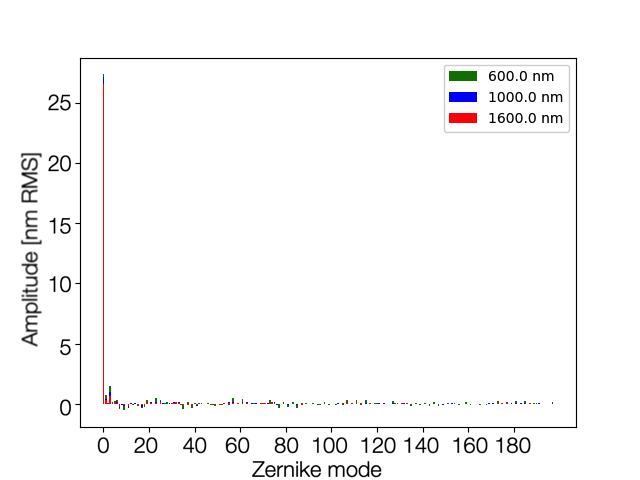}
        \caption{3PWFS}
    \end{subfigure}
    \hfill
    \begin{subfigure}[b]{0.32\textwidth}
        \centering
        \includegraphics[width=\textwidth]{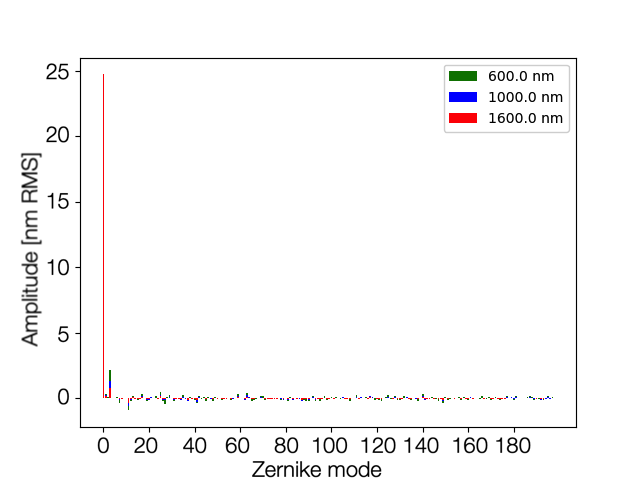}
        \caption{4PWFS}
    \end{subfigure}
    \hfill
    \begin{subfigure}[b]{0.32\textwidth}
        \centering
        \includegraphics[width=\textwidth]{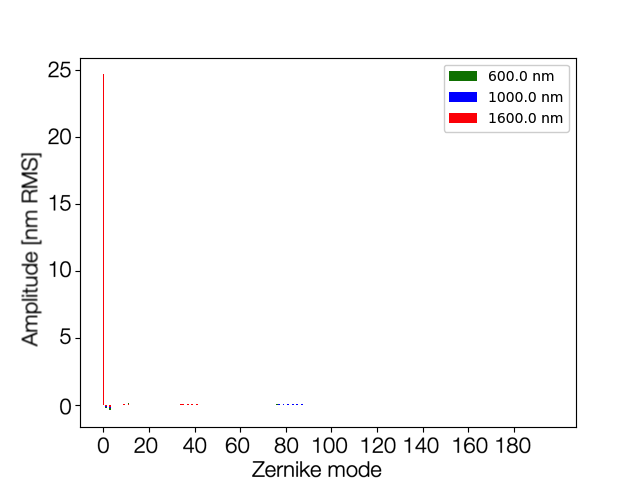}
        \caption{ZWFS}
    \end{subfigure}
    
    \caption{Phase measured by (a) 3PWFS, (b) 4PWFS, and (c) ZWFS when a petal mode with an amplitude of 25 nm RMS  is used as the input phase. The measurements are taken in open-loop and projected over the modified Zernike basis at wavelengths 0f 600nm, 1000nm, and 1600nm.}
    \label{fig:Open-loop_petal_mode}
\end{figure}

We use the petal mode, shown in Figure ~\ref{fig:petal_mode}, with an amplitude of 25 nm RMS as the input phase. IM for AO loop calibration is based on Zernike basis with first 200 zernike modes along with the petal mode and after removing the piston, tip and tilt modes. We call this basis the modified Zernike basis. We perform an open-loop AO simulation and compute the phase measured by the WFSs. The measured wavefront phase projected over the modified Zernike basis is shown in Figure ~\ref{fig:Open-loop_petal_mode}. We can see that the petal mode is accurately measured by all the three WFSs, with very little cross-talks with other modes. The cross-talks are slightly lesser at 1600 nm as compared to other two wavelenghts of 600 nm and 1000 nm. The cross-talks are the least for the ZWFS. 

\begin{figure}[H]
    \centering
    
    \begin{subfigure}[b]{0.32\textwidth}
        \centering
        \includegraphics[width=\textwidth]{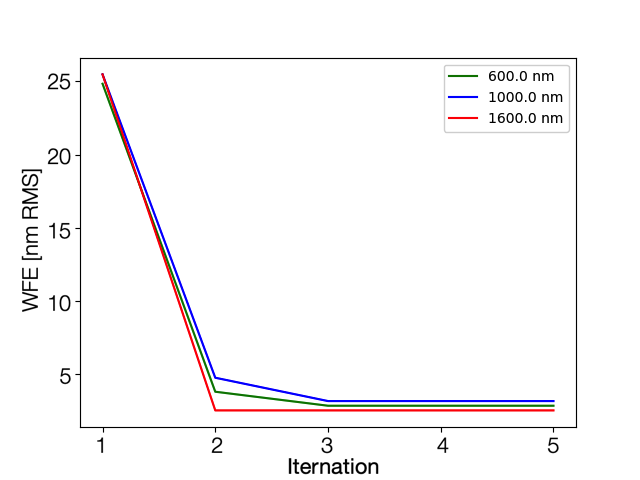}
        \caption{3PWFS}
    \end{subfigure}
    \hfill
    \begin{subfigure}[b]{0.32\textwidth}
        \centering
        \includegraphics[width=\textwidth]{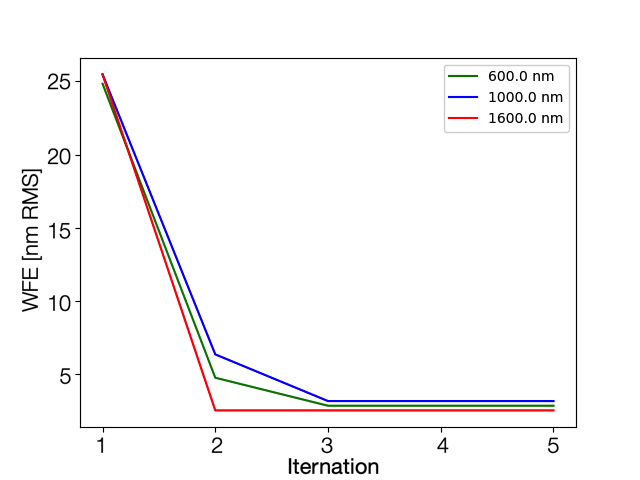}
        \caption{4PWFS}
    \end{subfigure}
    \hfill
    \begin{subfigure}[b]{0.32\textwidth}
        \centering
        \includegraphics[width=\textwidth]{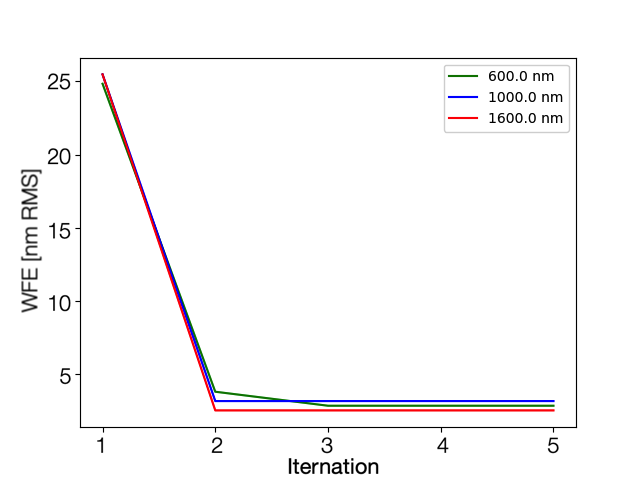}
        \caption{ZWFS}
    \end{subfigure}
    
    \caption{ Residual WFE progression with the number of AO loop iterations at wavelengths of 600 nm, 1000 nm, and 1600 nm, for (a) 3PWFS, (b) 4PWFS, and (c) ZWFS.}
    \label{fig:AO_closed-loop_residual_WFE}
\end{figure}

\begin{figure}[H]
    \centering
    
    \begin{subfigure}[b]{0.32\textwidth}
        \centering
        \includegraphics[width=\textwidth]{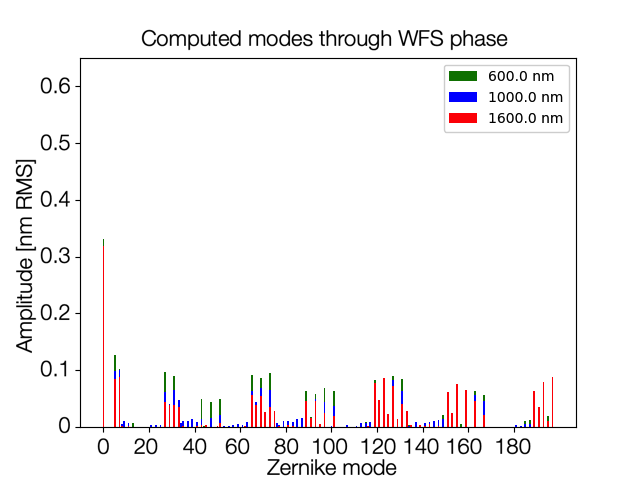}
    \end{subfigure}
    \hfill
    \begin{subfigure}[b]{0.32\textwidth}
        \centering
        \includegraphics[width=\textwidth]{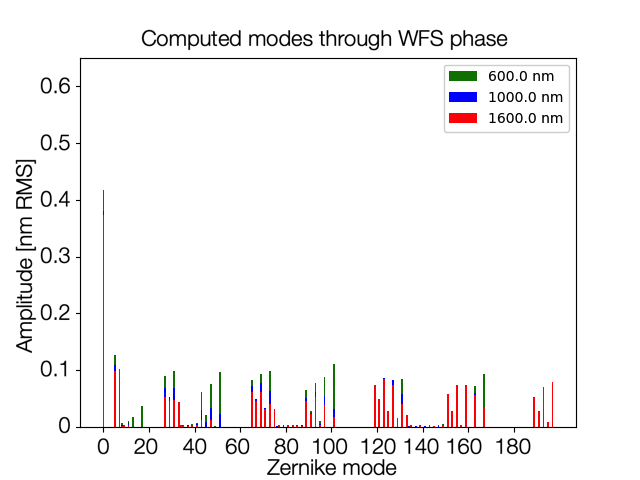}
    \end{subfigure}
    \hfill
    \begin{subfigure}[b]{0.32\textwidth}
        \centering
        \includegraphics[width=\textwidth]{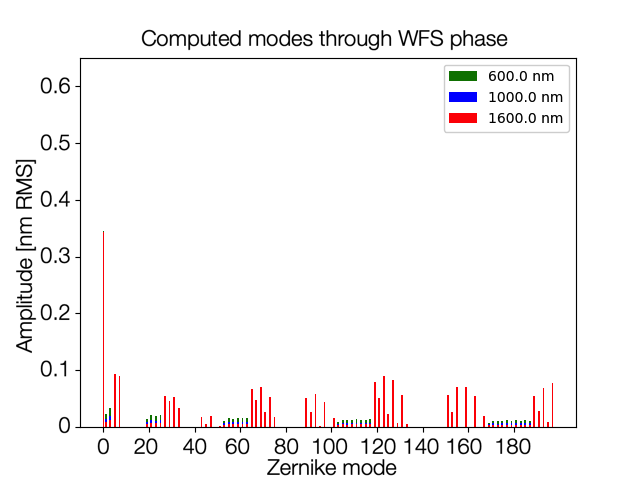}
    \end{subfigure}

    \begin{subfigure}[b]{0.32\textwidth}
        \centering
        \includegraphics[width=\textwidth]{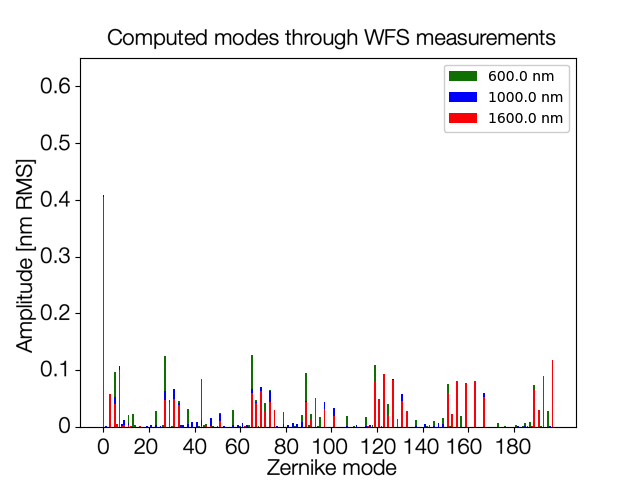}
    \end{subfigure}
    \hfill
    \begin{subfigure}[b]{0.32\textwidth}
        \centering
        \includegraphics[width=\textwidth]{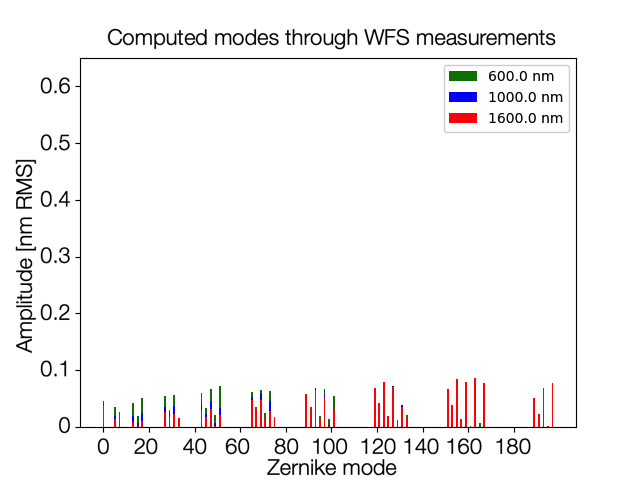}
    \end{subfigure}
    \hfill
    \begin{subfigure}[b]{0.32\textwidth}
        \centering
        \includegraphics[width=\textwidth]{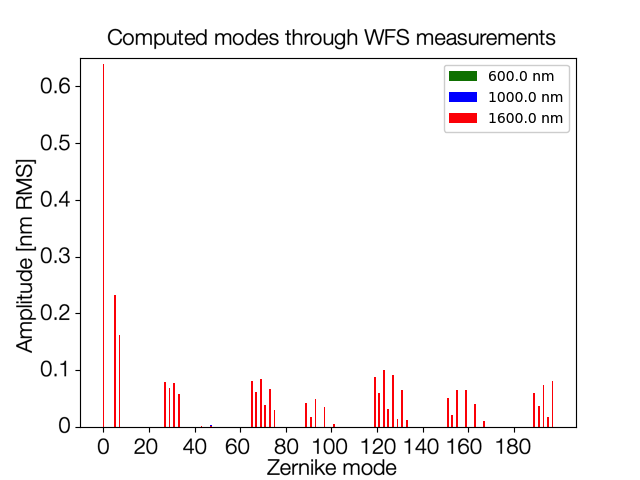}
    \end{subfigure}

    \begin{subfigure}[b]{0.32\textwidth}
        \centering
        \includegraphics[width=\textwidth]{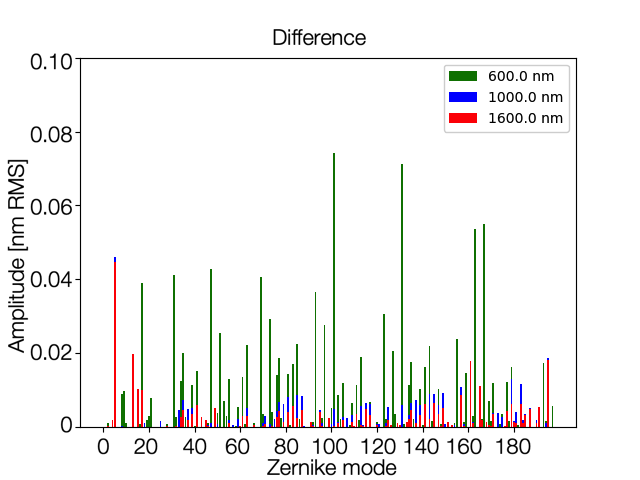}
        \caption{3PWFS}
    \end{subfigure}
    \hfill
    \begin{subfigure}[b]{0.32\textwidth}
        \centering
        \includegraphics[width=\textwidth]{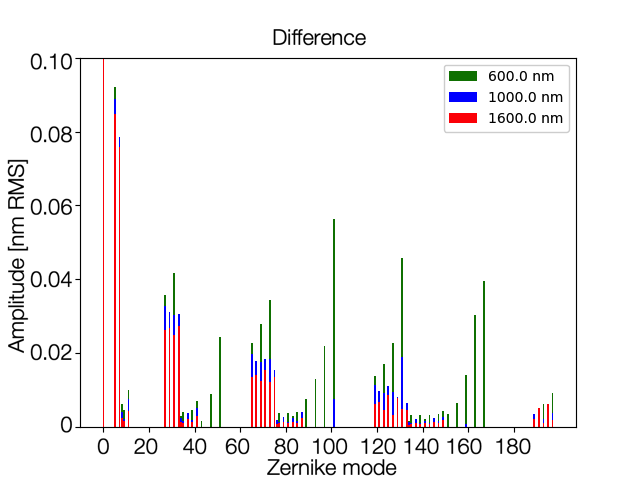}
        \caption{4PWFS}
    \end{subfigure}
    \hfill
    \begin{subfigure}[b]{0.32\textwidth}
        \centering
        \includegraphics[width=\textwidth]{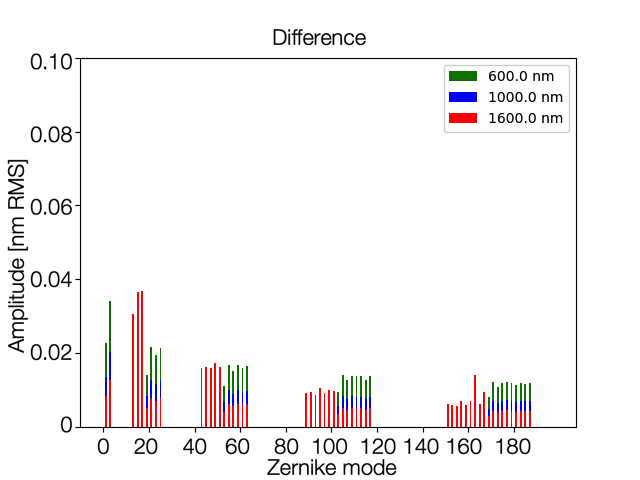}
        \caption{ZWFS}
    \end{subfigure}
    
    \caption{ Close-loop simulation results when petal mode with 25 nm RMS amplitude is used as the input phase. The residual phase is recorded after convergence of the loop (5 iterations) by (a) 3PWFS, (b) 4PWFS, and (c) ZWFS. The phase amplitude of modes in the modified Zernike basis is computed at the top through wavefront phase, in the middle through WFS measurements, and at the bottom by their difference, at wavelengths of 600 nm, 1000 nm, and 1600 nm.}
    \label{fig:Closed-loop_petal_mode}
\end{figure}

Further, we again use the petal mode with an amplitude of 25 nm RMS as the input phase and perform closed-loop AO simulations. The AO loop converges in 2-3 iterations, as shown in Figure ~\ref{fig:AO_closed-loop_residual_WFE}. The residual WFE is the least at 1600 nm for all the WFSs. At 600 nm, the residual WFE is higher because the amplitude of 25nm is at the limit of dynamic range. But at 1000nm, it is unclear
why the residual WFE is more than that at 600 nm. Figure ~\ref{fig:Closed-loop_petal_mode} shows the residual wavefront phase, the phase measured by the WFS and their difference projected over the modified Zernike basis, after the loop has converged. We can see that the residual wavefront phase is almost the same for all the WFSs and is limited by the DM’s inability to perfectly fit the discontinuity. We also observe that the ZWFS is best at measuring the residual phase, but the 4PWFS fails to correctly see the residual phase after the loop has converged, which could be a source of instability in practice. The 3PWFS can measure the residual phase quite accurately in this case. For both PWFS, a more thorough study is required, by changing relative angle of petal to pyramid (a parameter we may not be able to keep constant in practice). This explains why it takes longer for the AO loop to converge in case of the 4PWFS. Since the simulations performed at shorter wavelenghts show higher residuals once the loop is closed, which we think is due to higher non-linearities and modal confusion.

\section{Optical gains}

The non-linear characteristics of the FFWFSs induce a spatial frequency-dependent sensitivity loss during on-sky operations, which is quantified by optical gains. Calibration is performed using a point source and an IM, referred to as IMcalib, is obtained under ideal diffraction-limited conditions. However, during operation on-sky, we cannot attain the perfect diffraction limit of the telescope, causing deviations from the calibration regime in the WFS behavior. This non-linearity necessitates considering the WFS as a sensor with variable linear behavior, dependent upon the current atmospheric conditions. It is hypothesized that the on-sky behavior of the sensor can be described by another IM, termed IMonSky, as demonstrated in Chambouleyron et al. 2020 \cite{chambouleyron2020pyramid}.

The optical transfer matrix $T_{\text{opt}}$ describes the discrepancies between on-sky and calibration regimes. Considering the assumption by Deo et al. (2019) \cite{deo2018assessing} that there is no crosstalk between modes during the transition from calibration to on-sky regimes, $T_{\text{opt}}$ assumes a diagonal form. This allows us to express the modification of each mode's linear behavior by a scalar factor $G(\phi_i)$, denoted as the modal optical gain. The optical gain for each mode $\phi_i$ is computed as:

\begin{equation} \label{eqn}
G(\phi_i) = \frac{\langle \delta I_{\text{onSky}} (\phi_i) | \delta I_{\text{calib}} (\phi_i) \rangle}{\langle \delta I_{\text{calib}} (\phi_i) | \delta I_{\text{calib}} (\phi_i) \rangle}
\end{equation}

\noindent
An alternate formulation of this equation in terms of matrices is:

\begin{equation} \label{eqn}
G_{\text{opt}} = \frac{\text{diag}(\text{IM}_{\text{onSky}}^T \cdot \text{IM}_{\text{calib}})}{\text{diag}(\text{IM}_{\text{calib}}^T \cdot \text{IM}_{\text{calib}})}
\end{equation}

\noindent
where $G_{\text{opt}}$ is a vector containing all the $G(\phi_i)$ for $i$ in the range $[1, \text{N}]$.

\begin{figure}[H]
    \centering
    
    \begin{subfigure}[b]{0.84\textwidth}
        \centering
        \includegraphics[width=\textwidth]{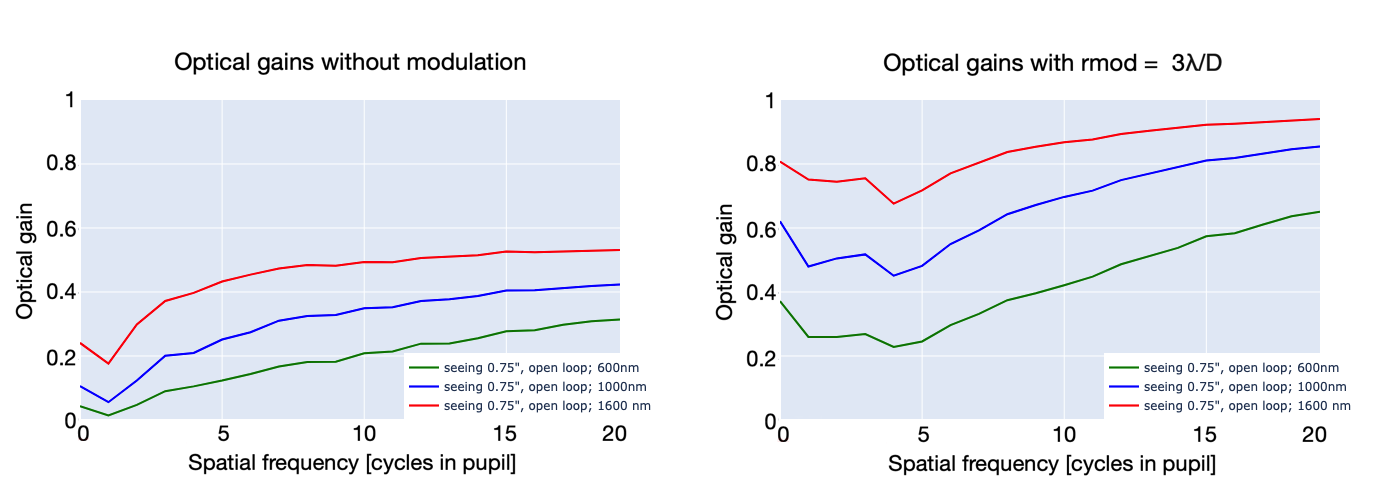}
        \caption{3PWFS}
    \end{subfigure}
    \hfill
    \begin{subfigure}[b]{0.84\textwidth}
        \centering
        \includegraphics[width=\textwidth]{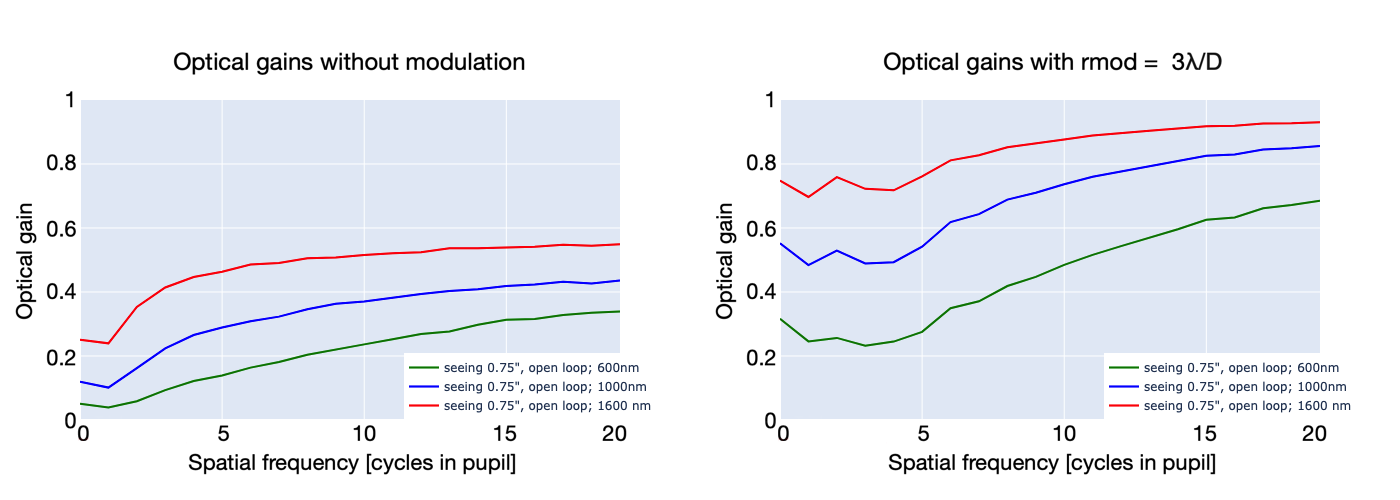}
        \caption{4PWFS}
    \end{subfigure}
    \hfill
    \begin{subfigure}[b]{0.42\textwidth}
        \centering
        \includegraphics[width=\textwidth]{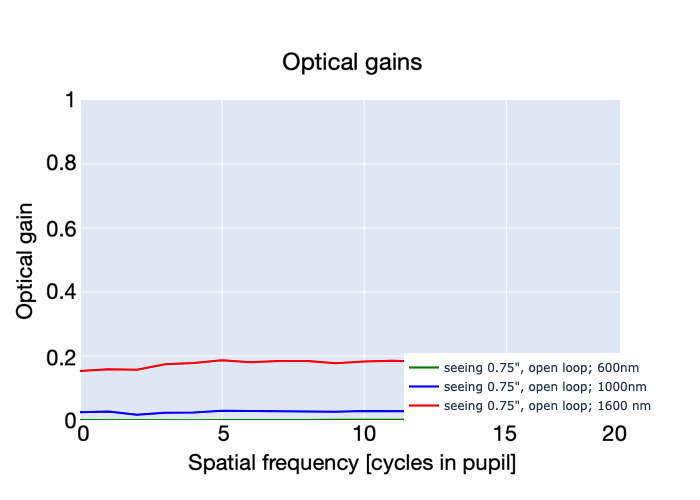}
        \caption{ZWFS}
    \end{subfigure}
    
    \caption{Open-loop optical gains as a function of spatial frequency plotted for the (a) 3PWFS, (b)4PWFS and (c) ZWFS, using phase screen generated after a open-loop AO simulation at seeing of 0.75 arcseconds.}
    \label{fig:Open-loop_optical_gains}
\end{figure}

We performed optical gain calculations for the 3PWFS, 4PWFS, and ZWFS at three distinct wavelengths under open-loop conditions, as shown in Figure ~\ref{fig:Open-loop_optical_gains}. The calibration matrix $\text{IM}_{\text{calib}}$ was obtained in the diffraction-limited regime. The on-sky calibration matrix $\text{IM}_{\text{onSky}}$ was obtained using independent phase screens at seeing of 0.75 arcseconds, representative of open and close loop situations with an XAO. Overall, optical gains increase with wavelength for all WFSs and are notably higher at 1600 nm. The 3PWFS exhibits optical gains similar to the 4PWFS in both the unmodulated case and with modulation of $3\lambda/D$.

In open-loop condition, ZWFS optical gains are nearly zero at 600 nm, indicating its inability to close a loop. Even at 1600 nm, optical gains for ZWFS remain very low. The optical gains in the unmodulated PWFS at 1600 nm are close to those of the PWFS with modulation of $3\lambda/D$ at 600 nm. This implies that going to 1600 nm has a similar effect to applying modulation, but without the loss of sensitivity for modes within the modulation radius.

\begin{figure}[H]
    \centering
    
    \begin{subfigure}[b]{0.84\textwidth}
        \centering
        \includegraphics[width=\textwidth]{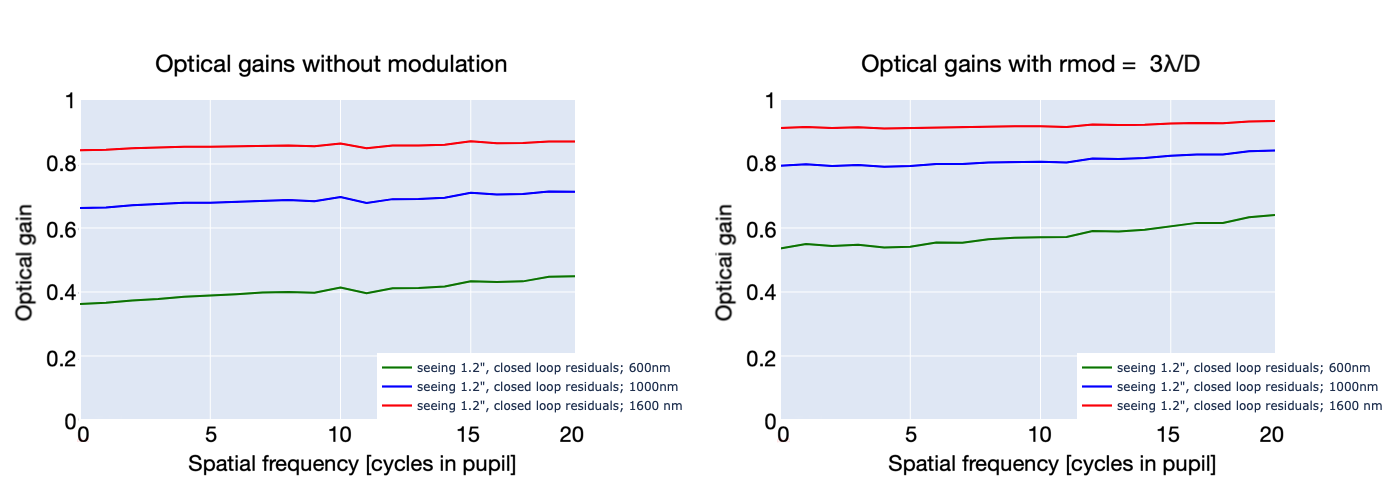}
        \caption{3PWFS}
    \end{subfigure}
    \hfill
    \begin{subfigure}[b]{0.84\textwidth}
        \centering
        \includegraphics[width=\textwidth]{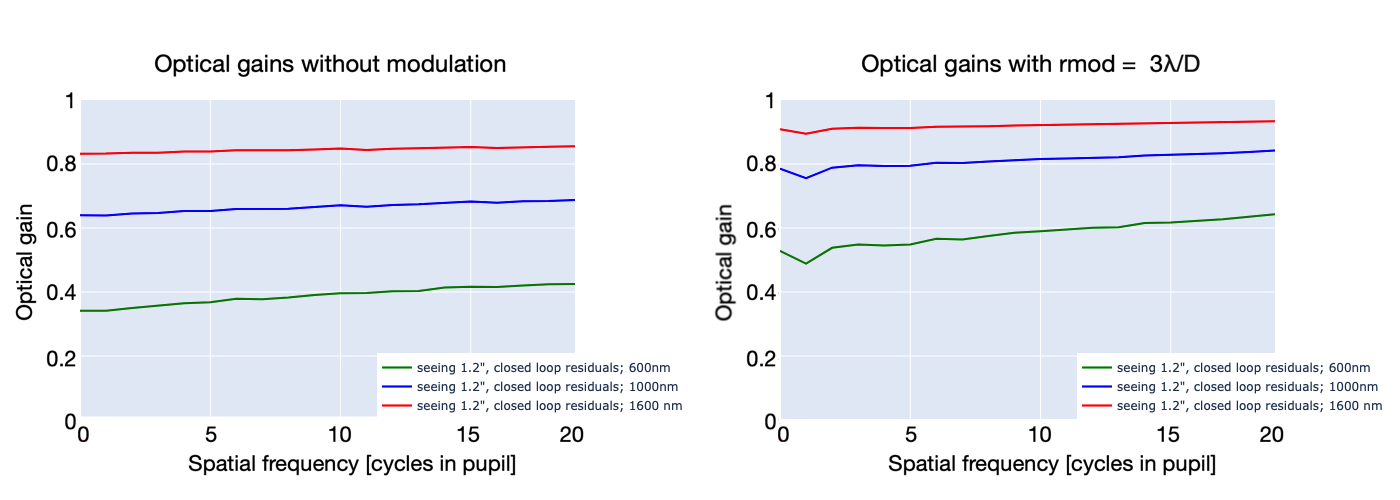}
        \caption{4PWFS}
    \end{subfigure}
    \hfill
    \begin{subfigure}[b]{0.42\textwidth}
        \centering
        \includegraphics[width=\textwidth]{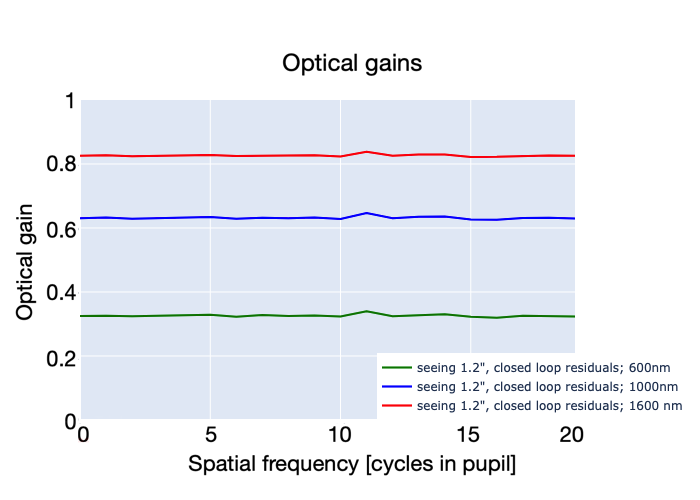}
        \caption{ZWFS}
    \end{subfigure}
    
    \caption{Closed-loop optical gains as a function of spatial frequency plotted for the (a) 3PWFs, (b)4PWFS and (c) ZWFS, using phase screen generated after a closed-loop AO simulation at seeing of 1.2 arcseconds.}
    \label{fig:Closed-loop_optical_gains}
\end{figure}

The optical gains under closed-loop conditions are shown in Figure ~\ref{fig:Closed-loop_optical_gains}. The calibration matrix $\text{IM}_{\text{calib}}$ was obtained in the diffraction-limited regime. The on-sky calibration matrix $\text{IM}_{\text{onSky}}$ was obtained by using a residual phase screen generated after a closed-loop AO simulation at seeing of 1.2 arcseconds. A higher seeing value is used for  closed-loop simulations as compared to open-loop simulations because we know that the residual phase would be quite low for closed-loop simulations. The 3PWFS and 4PWFS have similar optical gains in both the unmodulated case and with modulation of $3\lambda/D$. The ZWFS exhibits similar optical gains to the unmodulated PWFS. The optical gains fall flat in the closed-loop condition with a low residual phase, indicating that all modes are equally well controlled. Again, the optical gains are significantly higher at 1600 nm, which means that photons are better used in a FFWFS used at longer wavelengths. Also, modal confusion should be lowered and less variable from frame to frame. 

\section{Conclusion}

We demonstrated that going to longer wavelengths increases the dynamic range, optical gains and sensitivity for all FFWFSs. It also increases the ability of the WFS to sense petal modes with large amplitude.

In the transfer functions for the 3PWFS, we observed dead zones with minimal sensitivity in both the x and y transfer functions. However, these functions are complementary; their combined output can sense all spatial frequencies without any dead zones. Additionally, the definitions of x and y are related to the relative orientation of the pyramid to the detector, therefore these dead zones are not a concern. Rotating either the pyramid or the detector alters the x/y axis definition in relation to the pyramid. In fact the same is true for the 4PWFS.

The 3PWFS exhibits a marginally greater dynamic range compared to the 4PWFS for Zernike modes. Nonetheless, we noted that the 3PWFS presents an asymmetric response across zero to the petal mode under consideration. This asymmetry results from the 3PWFS's orientation relative to the petal mode. Conversely, the 4PWFS displays a symmetric response across zero to the petal mode. However, this symmetry is lost with more realistic petal shapes or when the petal's orientation differs relative to the 4PWFS. Overall, the 3PWFS's performance is comparable to, if not better than, the 4PWFS in terms of sensing petal modes during closed-loop simulations. It stands as a viable alternative, with a potential benefit on the manufacturing side. Our next steps involve conducting a similar analysis with more realistic VLT petal modes. We need to understand the effect of turbulence and low wind effects on the PWFS and ZWFS, particularly regarding their impact on optical gains, modal confusion, accuracy and, ultimately, loop stability.

Regarding optical gains, we noted that the ZWFS was unable to function in open-loop conditions at wavelengths of 600 and 1000 nm. Conversely, in closed-loop conditions, all three WFSs demonstrated comparable, stable optical gains.

We will soon start tests in the lab with unmodulated PWFS (and later ZWFS), focusing particularly on the lab demonstration of the 3PWFS for RISTRETTO, as well as the control of the low wind effect.

\acknowledgments     
 
This work has been carried out within the framework of the National Centre of Competence in Research PlanetS supported by the Swiss National Science Foundation under grants 51NF40\_182901 and 51NF40\_205606. The RISTRETTO project was partially funded through the SNSF FLARE programme for large infrastructures under grants 20FL21\_173604 and 20FL20\_186177. The authors acknowledge the financial support of the SNSF. The authors would also like to thank the RISTRETTO Front-end team for fruitful discussions.

\bibliographystyle{spiebib}
\bibliography{references}

\end{document}